\documentclass[%
 reprint,
superscriptaddress,
 amsmath,amssymb,
 aps,
]{revtex4-2}

\usepackage{graphicx}% Include figure files
\usepackage{dcolumn}% Align table columns on decimal point
\usepackage{bm}% bold math
\begin{document}

\preprint{APS/123-QED}

\title{Chiral phonons: circularly polarized Raman spectroscopy and \textit{ab initio} calculations in a chiral crystal tellurium}% Force line breaks with \\

\author{Kyosuke Ishito}
    \email{ishito.k.aa@m.titech.ac.jp}
    \affiliation{Department of Physics, Tokyo Institute of Technology, Tokyo 152-8551, Japan}%Lines break automatically or can be forced with \\
\author{Huiling Mao}
    \affiliation{Department of Physics, Tokyo Institute of Technology, Tokyo 152-8551, Japan}
\author{Kaya Kobayashi}
    \affiliation{Research Institute for Interdisciplinary Science, Okayama University, Okayama 700-8530, Japan}
\author{Yusuke Kousaka}
    \affiliation{Department of Physics and Electronics, Osaka Metropolitan University, Osaka 599-8531, Japan}
\author{Yoshihiko Togawa}
    \affiliation{Department of Physics and Electronics, Osaka Metropolitan University, Osaka 599-8531, Japan}
\author{Hiroaki Kusunose}
    \affiliation{Department of Physics, Meiji University, Kanagawa 214-8571, Japan}
\author{Jun-ichiro Kishine}
    \affiliation{Division of Natural and Environmental Sciences, The Open University of Japan, Chiba 261-8586, Japan}
\author{Takuya Satoh}
    \email{satoh@phys.titech.ac.jp}
    \affiliation{Department of Physics, Tokyo Institute of Technology, Tokyo 152-8551, Japan}

\date{\today}% It is always \today, today,

\begin{abstract}
Recently, phonons with chirality (chiral phonons) have attracted significant attention. Chiral phonons exhibit angular and pseudo-angular momenta. In circularly polarized Raman spectroscopy, the peak split of the $\Gamma_3$ mode is detectable along the principal axis of the chiral crystal in the backscattering configuration. In addition, peak splitting occurs when the  pseudo-angular momenta of the incident and scattered circularly polarized light are reversed. 
Until now, chiral phonons in binary crystals have been observed, whereas those in unary crystals have not been observed.
Here, we observe chiral phonons in a chiral unary crystal Te. 
The pseudo-angular momentum of the phonon is obtained in Te by an \textit{ab initio} calculation. From this calculation, we verified the conservation law of pseudo-angular momentum in Raman scattering.
From this conservation law, we determined the handedness of the chiral crystals. 
We also evaluated the true chirality of the phonons using a measure with symmetry similar to that of an electric toroidal monopole. 
%\begin{description}
%\end{description}
\end{abstract}

\maketitle

%\tableofcontents

\section{Introduction}

A phonon is a collective vibration of atoms in a crystal. In terms of a space group, chirality is the structural property wherein the object is different from its mirror image, and it plays a crucial role in optical physics. Optical activity is a well-known phenomenon derived from the chirality of a material. However, not only physical objects such as crystals and molecules have chirality but also dynamic motions such as phonons\cite{kishine20}. 
Recently, a chiral phonon with pseudo-angular (PAM)\cite{bozovicPossibleBandstructureShapes1984a,zhangChiralPhononsHighSymmetry2015} and angular momenta (AM)\cite{zhangAngularMomentumPhonons2014,garaninAngularMomentumSpinphonon2015,nakaneAngularMomentumPhonons2018} was proposed\cite{zhangChiralPhononsHighSymmetry2015} and observed\cite{zhuObservationChiralPhonons2018a}. The PAM of the phonon takes a quantized value originating from a phase change of a wave function acquired by a discrete rotation operation\cite{zhangChiralPhononsHighSymmetry2015} and is essential for conservation laws in light scattering \cite{damnjanovicSelectionRulesPolymers1983,tatsumiConservationLawAngular2018} and transport physics\cite{chen22}. 
The AM of a phonon has a continuous value calculated by the atomic circular motion and has been reported in binary crystals such as $\alpha$-HgS\cite{Ishito21} and ABi (A = K, Rb, Cs)\cite{Skorka22}, whereas that in unary crystals has not been reported.

Single-crystal Te is a unary chiral crystal. It has enantiomer structures with right-handed (RH) (P$3_121$) and left-handed helices (LH) (P$3_221$)\cite{Glazer86,furukawaCurrentinducedMagnetizationCaused2021}, which are denoted as R- and L-Te, respectively. 
From a macroscopic perspective, it exhibits various physical properties, such as a current-induced magnetization \cite{furukawaCurrentinducedMagnetizationCaused2021,furukawaObservationCurrentinducedBulk2017c,Tsirkin2018,Sahin2018}, electrical magnetochiral anisotropy \cite{rikkenStrongElectricalMagnetochiral2019}, sequential structural phase transitions under pressure \cite{ideuePressureinducedTopologicalPhase2019,Rodriguez2020,Akiba2020}, and optical activity \cite{nomuraOpticalActivityTellurium1960,Vorobev1979}. 
From a microscopic perspective, it has characteristic electron and phonon band structures.
Chiral atomic structure induces the linear band splitting \cite{bozovicPossibleBandstructureShapes1984a,teuchertSymmetryLatticeVibrations1974,Tsunetsugu22}
and radial spin texture\cite{sakanoRadialSpinTexture2020}. 
The phonon band of Te is calculated using various methods \cite{chen22,teuchertSymmetryLatticeVibrations1974,gibbonsCalculationThermalExpansion1973,orelLatticeDynamicsCrystalline1975,martinIntermolecularBondingLattice1976,ghoshOriginStabilityHelical2008,pengAnisotropicLatticeThermal2015,Hamada2018,Zhang22}.
The phonons of Te are measured by Raman spectroscopy \cite{torrieRamanSpectrumTellurium1970,pineRamanSpectraLattice1971,richterExtraordinaryPhononRaman1972,duOneDimensionalVanWaals2017,Jnawali2022} and infrared reflective spectroscopy \cite{keezerINFRAREDLATTICEBANDS1967}. 
However, the PAM and AM of the phonons in Te have not been discussed.

Here, we measured the phonons in Te using circularly polarized Raman spectroscopy and calculated the PAM and AM of the phonons using ABINIT (\textit{ab initio} calculation package) result. 
We also demonstrated that the conservation law of the PAM of the phonon determines the handedness of the chiral crystal and that the AM of the phonons in the chiral unary crystals becomes nonzero.

\section{material and method}
\subsection{Material}
We measured two samples of Te with $c$-axis-oriented surfaces.
Sample 1 was synthesized using the Bridgman method. Te powder of 5N was sealed in an evacuated quartz tube and cooled from 630$^\circ$C at a rate of 0.4 $\text{mm}/\text{h}$.
Sample 2 was purchased from MaTecK and had a purity of 6N.

\subsection{Raman spectroscopy}
\begin{figure*}
\includegraphics{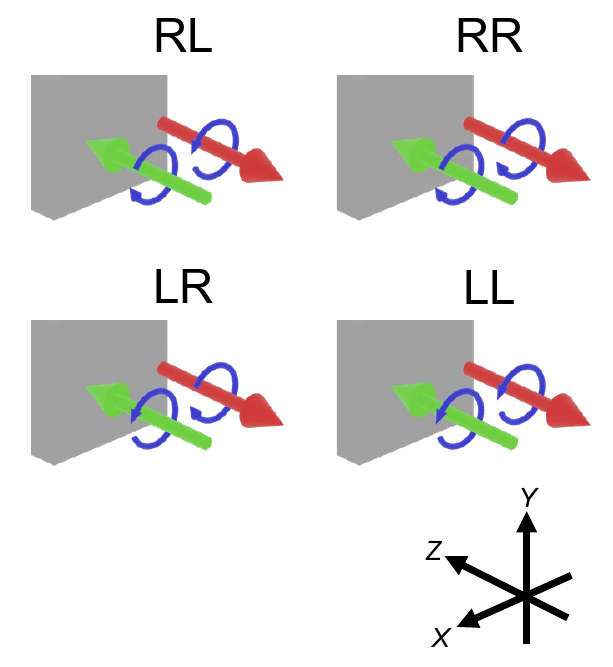}% Here is how to import EPS art
\caption{\label{fig:optical_system}
Backscattering geometry of circularly polarized Raman spectroscopy. 
The green and red arrows indicate incident and scattered light, respectively. The blue circular arrows indicate circular polarization, which is represented by R (right-handed) and L (left-handed). The wavelength of the incident light is $\lambda = 785 \, \text{nm}$.
}
\end{figure*}

We measured the phonons in Te samples using circularly polarized Raman spectroscopy at room temperature (RT).
Our setup for the polarization of the incident and scattered light is shown in Fig. \ref{fig:optical_system}.

We calculated the Raman selection rule based on the polarization of light and the Raman tensor.
The structure of Te belongs to the point group 32, and the Raman tensors are given as follows\cite{loudonRamanEffectCrystals1964}: 
\begin{gather}\label{A_mode}
\Gamma_1  : 
\begin{pmatrix}
  a & \cdot & \cdot \\
 \cdot  & a & \cdot \\
  \cdot & \cdot & b
  \end{pmatrix},\\
\Gamma_3(x) : 
\begin{pmatrix}
  c & \cdot & \cdot \\
  \cdot & -c & d \\
 \cdot  & d & \cdot
 \end{pmatrix},  \;\;\;
 \Gamma_3(y) : 
 \begin{pmatrix}
 \cdot  & -c & -d \\
 -c  & \cdot  & \cdot \\
 -d  & \cdot & \cdot
  \end{pmatrix}.
\end{gather}
Here, the polarization of light is expressed by the Jones vector $(1, i)/\sqrt{2}$ for right-handed (R) and $(1, -i)/\sqrt{2}$ for left-handed circular polarizations (L).

\subsection{\textit{Ab initio} calculation}
We calculated phonon dispersion in Te using density functional perturbation theory (DFPT)\cite{Hamann2005} in the ABINIT package\cite{Gonze2020,Romero2020}.
We used a projector augmented wave method \cite{Torrent2008,jolletGenerationProjectorAugmentedWave2014} as the pseudo-potential and generalized gradient approximation to the exchange-correlation potential in the Perdew–Burke–Ernzerhof form. Spin-orbit coupling was not considered in this study.
The parameters of the DFPT calculation were a 15 Ha energy cutoff and $8\times8\times6$ grids.
We used the lattice parameters of Te available in the material project \cite{Jain2013}: $a$ = $b$ = 4.512 \AA, $c$ = 5.960 \AA, $\alpha$ = $\beta$ = 90$^\circ$, $\gamma$ = 120$^\circ$ (Ref.\cite{osti_1193780}).

\section{Result}
\subsection{Raman spectroscopy}
In our condition, the incident and backscattered light, denoted as plane waves $e^{i(k_{\text{i}}z-\omega_{\text{i}} t)}$ and $e^{i(-k_{\text{s}}z-\omega_{\text{s}} t)}$, respectively, are parallel to the $c$ axis of the Te crystal. Thus, we detect a phonon with the wavenumber $k$ parallel to the $c$ axis. From the conservation law of the wavenumber, the phonon obtains a small but finite wavenumber $k$ in Raman scattering. The wavenumber is estimated as
\begin{eqnarray}
k = k_{\text{s}}+k_{\text{i}} \simeq 2k_{\text{i}} = 4\pi n/\lambda= 7.5 \times 10^{5} \ \text{rad}/\text{cm},
\end{eqnarray}
where $k_{\text{i}}$ and $k_{\text{s}}$ are the wavenumbers of the incident and scattered light, respectively, and absolute value of refractive index $n = 4.69$ (Ref. \cite{ciesielskiPermittivityGeTe2018}).
This wavenumber $k$ was 1.4\% that of A point in the Brillouin zone.
The Raman active phonon modes consist of $\Gamma_1+2\Gamma_3$ ($\Gamma_3^{(1)}$ and $\Gamma_3^{(2)}$) at $\Gamma$ point\cite{dresselhausGroupTheoryApplication2008a}.
We performed linearly polarized Raman spectroscopy to assign peaks to irreducible representations and listed the frequency of the phonons in Refs. \cite{torrieRamanSpectrumTellurium1970,pineRamanSpectraLattice1971} (Table 1).
Raman tensor calculation indicated the dependence of the polarization configuration.

We also performed circularly polarized Raman spectroscopy and measured each peak of the split $\Gamma_3$ mode in RL (R-incident and L-scattered) and LR (L-incident and R-scattered) configurations (Fig. 1). 
The phonon frequencies and their polarization dependencies for Sample 1 and 2 are shown in Figs. 2 and 3, respectively.
Two $\Gamma_3$ mode peaks split, and the phonon frequency of RL peaks was higher than that of LR peaks in Sample 1. The two $\Gamma_3$ mode peaks split, and the phonon frequency of the LR peaks was higher than that of the RL peaks in Sample 2.
The frequency differences between the RL and LR peaks of Sample 1 and 2 and those in Ref. \cite{pineRamanSpectraLattice1971} are listed in Table 2.
Such splitting has also been observed in Raman spectroscopy in $\alpha$-quartz\cite{pineLinearWaveVectorShifts1969,Oishi22}, Te\cite{pineRamanSpectraLattice1971}, and $\alpha$-HgS\cite{Ishito21}.

\subsection{\textit{Ab initio} calculation and the AM analysis of the phonon}
We calculated the phonon dispersion of R- and L-Te (Figs. 4(a) and (e), respectively ), which is consistent with previous \textit{ab initio} calculations \cite{chen22,ghoshOriginStabilityHelical2008,pengAnisotropicLatticeThermal2015,Hamada2018}.
The phonon frequencies at the $\Gamma$ point of the \textit{ab initio} calculation matched those of Raman spectroscopy within 10 $\text{cm}^{-1}$. 
The AM and PAM of the phonons were assigned to the phonon dispersion curves (Figs. 4 and 5), respectively, and they were consistent with Ref. \cite{chen22}.

The $z$ component of the AM of the phonon $m_{z,\text{AM}}$ of branch $j$ at wavevector $\bf{k}$ is defined as\cite{zhangAngularMomentumPhonons2014}
\begin{eqnarray}
m_{z,\text{AM}}(\textbf{k}, j)=\left(\boldsymbol \epsilon (\textbf{k}, j) ^{\dagger }M_z\boldsymbol \epsilon (\textbf{k}, j) \right) \hbar,
\end{eqnarray}
where $\boldsymbol{\epsilon} (\textbf{k}, j)$ is the eigenvector of the dynamical matrix
\begin{eqnarray}
M_z = \begin{pmatrix}
				 0 & -i & 0 \\
				 i & 0 & 0 \\
				 0 & 0 & 0 \\
\end{pmatrix}
\otimes I_{n\times n},
\end{eqnarray}
where the basis of the $3\times 3$ matrices is represented by the orthogonal bases $(u_x,u_y,u_z)$, $n$ is the number of atoms in a unit cell, and $I_{n\times n}$ is a unit matrix of $n\times n$. 

The PAM of the phonon $m_{\text{PAM}}$ of branch $j$ at wavevector $\bf{k}$ in $C_3$ invariant material is defined as\cite{zhangChiralPhononsHighSymmetry2015}
\begin{eqnarray}
\left\{C_3~|~0\right\} \textbf{u}(\textbf{k})=\exp \left[-i\frac{2\pi}{3}
	m_{\text{PAM}}(\textbf{k})\right]\textbf{u}(\textbf{k}),
\end{eqnarray}
where $\left\{C_3~|~0\right\}$ indicates $C_3$ rotation around the $c$ axis with no translation, and $\textbf{u}(\textbf{k})$ is the phonon displacement at wavenumber $\bf{k}$. 
However, the symmetry operation of Te involves helical rotation.
For example, R-Te is $\left\{C_3~|~1/3\right\}$ invariant, where $\left\{C_3~|~1/3\right\}$ indicates $C_3$ rotation around the $c$ axis and $c/3$ translation along the $c$ axis. 
Therefore, the spin PAM of phonon $m_{\text{PAM}}^{\text{s}}(\text{RH}, \bf{k})$ of branch $j$ at wavenumber $\bf{k}$ in R-Te is defined as \cite{Ishito21,Zhang22} 
\begin{eqnarray}
&\left\{C_3~|~\mathbf{c}/3\right\} \textbf{u}(\text{RH}, \textbf{k})&  \nonumber \\
&=\exp \left[-i\frac{2\pi}{3} 
	m_{\text{PAM}}^{\text{s}}(\text{RH}, \textbf{k})+\frac{\bf{k}\cdot \bf{c}}{3} \right]\textbf{u}(\text{RH}, \textbf{k}).&
\end{eqnarray}
Similarly, the spin PAM of the phonon $m_{\text{PAM}}^{\text{s}}(\text{LH}, \bf{k})$ of branch $j$ at wavenumber $\bf{k}$ in L-Te is defined as \cite{Ishito21,Zhang22} 
\begin{eqnarray}
&\left\{C_3~|~ 2\mathbf{c}/3\right\} \textbf{u}(\text{LH}, \textbf{k})&  \nonumber \\
&=\exp \left[-i\frac{2\pi}{3} 
	m_{\text{PAM}}^{\text{s}}(\text{LH}, \textbf{k})+\frac{2\bf{k}\cdot \bf{c}}{3} \right]\textbf{u}(\text{LH}, \textbf{k}).&
\end{eqnarray}
Figure 4 shows that there are nonzero AM parts in the phonon dispersion of Te. 
From Figs. 4 and 5, double-degenerated $\Gamma_3$ phonons split along $\Gamma$ to A points. 
Moreover, the signs of $m_{\text{AM}}$ and $m_{\text{PAM}}^{\text{s}}$ along the $\Gamma$ to A point reversed, and their absolute values were invariant after changing the handedness of Te.

We also characterized the true chirality of the phonons quantitatively by evaluating a pseudoscalar quantity similar to that of the electric toroidal monopole, $G_0$\cite{Oiwa22,Yamamoto22}.
To this end, we introduced the structure factors, $f_{x}(\textbf{k})=s_{1}-s_{2}$, $f_{y}(\textbf{k})=(s_{1}+s_{2}-2s_{3})/\sqrt{3}$, $f_{z}(\textbf{k})=(2/\sqrt{6})(s_{1}+s_{2}+s_{3})$, where $s_{i}=\sin(\textbf{k}\cdot\boldsymbol{\eta}_{i})$ with the nearest-neighbor bonds, $\boldsymbol{\eta}_{1}=(3\zeta a/2,\sqrt{3} \zeta a/2,c/3)$, $\boldsymbol{\eta}_{2}=(-3 \zeta a/2,\sqrt{3} \zeta a/2,c/3)$, $\boldsymbol{\eta}_{3}=(0,-\sqrt{3} \zeta a,c/3)$ ($\zeta \simeq 0.274$).
Note that $\textbf{f}(\textbf{k})=(f_{x},f_{y},f_{z})$ transforms like a polar vector and is proportional to $(k_{x},k_{y},k_{z})$ in the limit of $\textbf{k}\to0$.
Then, the true chirality of the phonons is evaluated by the dimensionless quantities defined as
\begin{align}
G_0 = \frac{1}{N\hbar}\sum_{\textbf{k}j}^{\rm BZ}\biggl\{&\textbf{m}_{{\rm AM}}(\textbf{k},j)\cdot\textbf{f}(\textbf{k})\biggr\}n\bigl(\omega(\textbf{k},j)\bigr),
\cr
G_u = \frac{1}{N\hbar}\sum_{\textbf{k}j}^{\rm BZ}\biggl\{&3m_{z,{\rm AM}}(\textbf{k},j)f_{z}(\textbf{k})\cr
&-\textbf{m}_{{\rm AM}}(\textbf{k},j)\cdot\textbf{f}(\textbf{k})\biggr\}n\bigl(\omega(\textbf{k},j)\bigr),
\end{align}
where $\omega(\textbf{k},j)$ is the phonon frequency of branch $j$ and the summation is taken over all branches and the Brillouin zone (BZ).
$N$ is the number of lattice sites, and $n(\omega) = 1/(e^{\hbar \omega/k_{\text{B}}T}-1)$ is the Bose-Einstein distribution function at the temperature $T$.
The chiralization and the handedness of the phonons are characterized by the magnitude and the sign of $G_{0}$, while $G_u$ characterizes the anisotropy of true chirality between the longitudinal and perpendicular directions to the helical axis.
$G_{0} \simeq 1 \times 10^{-1}$ and $ -1 \times 10^{-1}$ for L-Te and R-Te at $300$ K, respectively, 
and $G_u \simeq 3 \times 10^{-1}$ and $ -3 \times 10^{-1}$ for L-Te and R-Te at $300$ K, respectively.
The positive $G_u/G_0$ indicates the importance of the longitudinal contributions.
The signs of $G_0$ and $G_u$ are opposite with each other for a pair of enantiomers of Te.

\section{discussion}
The Raman selection rule is determined using a Raman tensor derived from the crystal point group\cite{loudonRamanEffectCrystals1964}. 
Considering the Raman tensor provides a selection rule for the phonon at the $\Gamma$ point
\begin{eqnarray}
\Gamma_1 \text{ mode: }I_{\text{RR}}:I_{\text{LL}}:I_{\text{RL}}:I_{\text{LR}} = 1 : 1 : 0 : 0,\\
\Gamma_3 \text{ mode: }I_{\text{RR}}:I_{\text{LL}}:I_{\text{RL}}:I_{\text{LR}} = 0 : 0 : 1 : 1, 
\end{eqnarray}
where the incident and scattered light are parallel to the $c$ axis.
Theoretically, in the RL and LR configurations, peaks of the $\Gamma_1$ mode are absent.
However, in our experiment, we observed the $\Gamma_1$ mode.
There are two possible reasons for this.
First, the incident light was not parallel to the $c$ axis.
Because we used an objective lens to focus plane wave light on the sample, part of the incident light was tilted.
Second, the surfaces of the samples were not sufficiently smooth.
Some parts of the surface were tilted from the $c$-axis-oriented surface because of the rough surface. 

We verified the spin PAM conservation law\cite{Ishito21} for the Raman scattering process as follows:
\begin{eqnarray}
    \sigma_{\text{i}} - \sigma_{\text{s}} \equiv \pm m_{\text{PAM}}^{\text{s}} \, (\text{mod} \, 3), \label{conservation_law}
\end{eqnarray}
where $\sigma_{\text{i}}$ and $\sigma_{\text{s}}$ represent the spin PAM of the incident and scattered photons, respectively. Furthermore, plus and minus correspond to the Stokes and anti-Stokes processes, respectively, on the right-hand side of Eq. (\ref{conservation_law}) (Ref. \cite{tatsumiConservationLawAngular2018}), respectively.
The R-photon and L-photon have spin PAM $\sigma=+1$ and $-1$, respectively\cite{Yariv06}. 
Through the Raman scattering process, the phonon obtains the spin PAM from the change in the spin PAM of photons.
In the RL configuration, a change in the spin PAM of photons $+1-(-1) \equiv -1 \, (\text{mod} \, 3)$ is transferred to phonons.
However, in the LR configuration, a change in the spin PAM of photons $-1-(+1) \equiv +1 \, (\text{mod} \, 3)$ is transferred to phonons.
In circularly polarized Raman spectroscopy of Sample 1, a higher peak of the split $\Gamma_3$ mode is observed in the RL configuration and corresponds to $m_{\text{PAM}}^{\text{s}} = -1$. 
Indeed, in R-Te, the phonon of the split $\Gamma_3$ mode with a higher frequency has $m_{\text{PAM}}^{\text{s}} = -1$ whereas in L-Te, the phonon of the split $\Gamma_3$ mode with a higher frequency has $ m_{\text{PAM}}^{\text{s}} = +1$.
Therefore, we presumed that Sample 1 and 2 were R- and L-Te, respectively.
Generally, because the lack of all mirror and inversion symmetries induces $\Gamma_3$ mode splitting, chiral materials exhibit $\Gamma_3$ mode splitting \cite{bozovicPossibleBandstructureShapes1984a,Tsunetsugu22}.
A combination of $\Gamma_3$ mode splitting in chiral materials and spin PAM assignment enables us to determine the handedness of other chiral materials, such as $\alpha$-quartz and $\alpha$-HgS\cite{Ishito21}.

\section{conclusion}
Circularly polarized Raman spectroscopy provided phonon frequencies of $\Gamma_1$ and $\Gamma_3$ modes in unary chiral crystal Te.
The splitting of the $\Gamma_3$ mode was reversed in the RL and LR configurations.  
These results can be explained by the spin PAM conservation law through the Raman scattering process.
This process determines the handedness of chiral crystals.
Moreover, we showed that the PAM, AM, and $G_0$ of a phonon in a chiral unary crystal Te are nonzero.
This indicates that the phonons in Te have true chirality.
The conversion of AM has been extended to the spin polarization of electrons in various chiral materials\cite{Ohe}.

\section*{acknowledgements}
We thank K. Matsumoto, T. Zhang, S. Murakami, H. Matsuura, and H. M. Yamamoto for their valuable discussions and technical support. 
This work was supported financially by the Japan Society for the Promotion of Science KAKENHI (grant nos. JP19H01828, JP19H05618, JP19K21854, JP21H01031, JP21H01032, and JP22H01154), the Frontier Photonic Sciences Project of the National Institutes of Natural Sciences (grant nos. 01212002 and 01213004), and MEXT Initiative to Establish NeXt-generation Novel Integrated Circuits CenterS (X-NICS) (grant no. JPJ011438).

%\section*{conflict of interest}
%The authors declare that they have no conflict of interest.

%\section*{Supporting Information}

\printendnotes

% Submissions are not required to reflect the precise reference formatting of the journal (use of italics, bold etc.), however it is important that all key elements of each reference are included.
%\bibliography{sample}

\begin{thebibliography}{99}
\bibitem{kishine20}
Kishine J, Ovchinnikov AS, Tereshchenko AA.
Chirality-induced phonon dispersion in a noncentrosymmetric micropolar crystal. 
\textit{Phys Rev Lett} \textbf{2020};125(24):245302.

\bibitem{bozovicPossibleBandstructureShapes1984a}
Bo\ifmmode \check{z}\else \v{z}\fi{}ovic I.
Possible band-structure shapes of quasi-one-dimensional solids.
\textit{Phys Rev B} \textbf{1984};29(12):6586--6599.

\bibitem{zhangChiralPhononsHighSymmetry2015}
Zhang L, Niu Q. 
Chiral Phonons at High-Symmetry Points in Monolayer Hexagonal Lattices. 
\textit{Phys Rev Lett} \textbf{2015};115(11):115502.


\bibitem{zhangAngularMomentumPhonons2014}
Zhang L, Niu Q.
Angular Momentum of Phonons and the Einstein--de Haas Effect.
\textit{Phys Rev Lett} \textbf{2014};112(8):085503.

\bibitem{garaninAngularMomentumSpinphonon2015}
Garanin DA, Chudnovsky EM.
Angular momentum in spin-phonon processes.
\textit{Phys Rev B} \textbf{2015};92(2):024421.

\bibitem{nakaneAngularMomentumPhonons2018}
Nakane JJ, Kohno H.
Angular momentum of phonons and its application to single-spin relaxation.
\textit{Phys Rev B} \textbf{2018};97(17):174403.

\bibitem{zhuObservationChiralPhonons2018a}
Zhu H, Yi J, Li M-Y, et al.
Observation of chiral phonons.
\textit{Science} \textbf{2018};359(6375):579--582.

\bibitem{damnjanovicSelectionRulesPolymers1983}
Damnjanovic M, Bozovic I, Bozovic N.
Selection rules for polymers and quasi-one-dimensional crystals. I. Kronecker products for the line groups isogonal to $\text{C}_{\text{n}}$, $\text{C}_{\text{nv}}$, $\text{C}_{\text{nh}}$ and $\text{S}_{\text{2n}}$
\textit{J Phys A Math Gen} \textbf{1982};16(17):3937--3947.

%\bibitem{chamberlainTheoryOnephononRaman1995a}
%Chamberlain MP, Trallero-Giner C, Cardona M.
%Theory of one-phonon Raman scattering in semiconductor microcrystallites.
%\textit{Phys Rev B} \textbf{1995};51(3):1680--1693.

\bibitem{tatsumiConservationLawAngular2018}
Tatsumi Y, Kaneko T, Saito R.
Conservation law of angular momentum in helicity-dependent Raman and Rayleigh scattering.
\textit{Phys Rev B} \textbf{2018};97(19):195444.

\bibitem{chen22}
Chen H, Wu W, Zhu J, et al.
Chiral Phonon Diode Effect in Chiral Crystals.
\textit{Nano Lett} \textbf{2022};22(4):1688--1693.

\bibitem{Ishito21}
%\textcolor{green}{
%Truly chiral phonons in $\alpha$-HgS observed by circularly polarised Raman spectroscopy
%K. Ishito, H. Mao, Y. Kousaka, Y. Togawa, S. Iwasaki, T. Zhang, S. Murakami, J. Kishine, and T. Satoh, arXiv , 2110.11604 (2021).
%}

Ishito K, Mao H, Kousaka Y, et al.
Truly chiral phonons in $\alpha$-HgS.
%\textit{arXiv preprint} \textbf{2021};arXiv:2110.11604 [cond-mat.mtrl-sci].
\textit{Nat Phys} \textbf{2023};19(1):35--39.


\bibitem{Skorka22}
%\textcolor{green}{
%J. Sk\'orka, K. J. Kapcia, P. T. Jochym and A. Ptok, arXiv:2203.05524v1 [cond-mat.mtrl-sci] (2022)
%}
Sk\'orka J, Kapcia KJ, Jochym PT, Ptok A.
Chiral phonons in binary compounds $A$Bi ($A$ = K, Rb, Cs) with P$2_1$/c structure.
\textit{arXiv preprint} \textbf{2022};arXiv:2203.05524.


\bibitem{Glazer86}
Glazer AM, Stadnicka K.
On the origin of optical activity in crystal structures.
\textit{J Appl Crystallogr} \textbf{1986};19(2):108--122.

\bibitem{furukawaCurrentinducedMagnetizationCaused2021}
Furukawa T, Watanabe Y, Ogasawara N, Kobayashi K, Itou T.
Current-induced magnetization caused by crystal chirality in nonmagnetic elemental tellurium.
\textit{Phys Rev Res} \textbf{2021};3(2):023111.

\bibitem{furukawaObservationCurrentinducedBulk2017c}
Furukawa T, Shimokawa Y, Kobayashi K, Itou T.
Observation of current-induced bulk magnetization in elemental tellurium.
\textit{Nat Commun} \textbf{2017};8(1):954.

\bibitem{Tsirkin2018}
Tsirkin SS, Puente PA, Souza I.
Gyrotropic effects in trigonal tellurium studied from first principles.
\textit{Phys Rev B} \textbf{2018};97(3):035158.

\bibitem{Sahin2018}
\ifmmode \mbox{\c{S}}\else \c{S}\fi{}ahin C, Rou J, Ma J, Pesin DA.
Pancharatnam-Berry phase and kinetic magnetoelectric effect in trigonal tellurium.
\textit{Phys Rev B} \textbf{2018};97(20):205206.

\bibitem{rikkenStrongElectricalMagnetochiral2019}
Rikken GLJA, Avarvari N.
Strong electrical magnetochiral anisotropy in tellurium.
\textit{Phys Rev B} \textbf{2019};99(24):245153.

%\bibitem{akahamaPressureinducedSuperconductivityPhase1992}
%Akahama Y, Kobayashi M, Kawamura H.
%Pressure-Induced Superconductivity and Phase Transition in Selenium and Tellurium.
%\textit{Solid State Commun} \textbf{1992};84(8):803--806.


\bibitem{ideuePressureinducedTopologicalPhase2019}
Ideue T, Hirayama M, Taiko H, et al.
Pressure-Induced Topological Phase Transition in Noncentrosymmetric Elemental Tellurium.
\textit{Proc Natl Acad Sci USA} \textbf{2019};116(51):25530--25534.

%\bibitem{rodriguezOpticalSignaturesPhase2020}
%Rodriguez D, Tsirlin AA, Biesner T, et al.
%Optical Signatures of Phase Transitions and Structural Modulation in Elemental Tellurium under Pressure.
%\textit{Phys Rev B} \textbf{2020}a;101(17):174104.

\bibitem{Rodriguez2020}
Rodriguez D, Tsirlin AA, Biesner T, et al.
Two Linear Regimes in Optical Conductivity of a Type-I Weyl Semimetal: The Case of Elemental Tellurium.
\textit{Phys Rev Lett} \textbf{2020};124(13):136402.

\bibitem{Akiba2020}
Akiba K, Kobayashi K, Kobayashi TC, et al.
Magnetotransport properties of tellurium under extreme conditions.
\textit{Phys Rev B}  \textbf{2020};101(24):245111.


\bibitem{nomuraOpticalActivityTellurium1960}
Nomura KC.
Optical Activity in Tellurium.
\textit{Phys Rev Lett} \textbf{1960};5(11):500--501.

\bibitem{Vorobev1979}
{Vorob'ev} LE, {Ivchenko} EL, {Pikus} GE, et al.
Optical activity in tellurium induced by a current.
\textit{JETP Lett} \textbf{1979};29:441.

\bibitem{teuchertSymmetryLatticeVibrations1974}
Teuchert WD, Geick R.
Symmetry of Lattice Vibrations in Selenium and Tellurium.
\textit{Phys Status Solidi B} \textbf{1974};61(1):123--136.

\bibitem{Tsunetsugu22}
Tsunetsugu H, Kusunose H.
Theory of Energy Dispersion of Chiral Phonons.
\textit{J Phys Soc Jpn};\textbf{2023};92(1):023601.

\bibitem{sakanoRadialSpinTexture2020}

Sakano M, Hirayama M, Takahashi T, et al.
Radial Spin Texture in Elemental Tellurium with Chiral Crystal Structure.
\textit{Phys Rev Lett} \textbf{2020};124(13):136404.

\bibitem{gibbonsCalculationThermalExpansion1973}
Gibbons TG.
Calculation of the Thermal Expansion for a Quasiharmonic Model of Tellurium.
\textit{Phys Rev B} \textbf{1973};7(4):1410--1419.

\bibitem{orelLatticeDynamicsCrystalline1975}
Orel B, Tubino R, Zerbi G.
Lattice dynamics of crystalline tellurium.
\textit{Mol Phys} \textbf{1975};30(1):37--48.

\bibitem{martinIntermolecularBondingLattice1976}
Martin RM, Lucovsky G, Helliwell K.
Intermolecular bonding and lattice dynamics of Se and Te.
\textit{Phys Rev B} \textbf{1976};13(4):1383--1395.

\bibitem{ghoshOriginStabilityHelical2008}
Ghosh P, Bhattacharjee J, Waghmare UV.
The Origin of Stability of Helical Structure of Tellurium.
\textit{J Phys Chem C} \textbf{2008};112(4):983--989.

\bibitem{pengAnisotropicLatticeThermal2015}
Peng H, Kioussis N, Stewart DA.
Anisotropic lattice thermal conductivity in chiral tellurium from first principles.
\textit{Appl Phys Lett} \textbf{2015};107(25):251904.

\bibitem{Hamada2018}
Hamada M, Minamitani E, Hirayama M, Murakami S.
Phonon Angular Momentum Induced by the Temperature Gradient.
\textit{Phys Rev Lett} \textbf{2018};121(17):175301.

\bibitem{Zhang22}
Zhang T, Murakami S, 
Chiral phonons and pseudoangular momentum in nonsymmorphic systems.
\textit{Phys Rev Res} \textbf{2022};4(1):L012024.

\bibitem{torrieRamanSpectrumTellurium1970}
Torrie BH.
Raman spectrum of tellurium.
\textit{Solid State Commun} \textbf{1970};8(22):1899--1901.

\bibitem{pineRamanSpectraLattice1971}
Pine AS, Dresselhaus G.
Raman Spectra and Lattice Dynamics of Tellurium.
\textit{Phys Rev B} \textbf{1971};4(2):356--371.

\bibitem{richterExtraordinaryPhononRaman1972}
Richter W.
Extraordinary phonon Raman scattering and resonance enhancement in tellurium.
\textit{J Phys Chem Solids} \textbf{1972};33(11):2123--2128.

\bibitem{duOneDimensionalVanWaals2017}
Du Y, Qiu G, Wang Y, et al.
One-Dimensional van der Waals Material Tellurium: Raman Spectroscopy under Strain and Magneto-Transport.
\textit{Nano Lett} \textbf{2017};17(6):3965--3973.

\bibitem{Jnawali2022}
Jnawali G, Xiang Y, Linser SM, et al. 
Ultrafast photoinduced band splitting and carrier dynamics in chiral tellurium nanosheets.
\textit{Nat Commun} \textbf{2020};11:3991.

\bibitem{keezerINFRAREDLATTICEBANDS1967}
Lucovsky G, Keezer RC, Burstein E.
Infra-red lattice bands of trigonal tellurium and selenium.
\textit{Solid State Commun} \textbf{1967};5(6):439--445.

\bibitem{loudonRamanEffectCrystals1964}
Loudon R.
The Raman effect in crystals.
\textit{Adv Phys} \textbf{1964};13(52):423--482.

\bibitem{Hamann2005}
Hamann DR, Wu X, Rabe KM, Vanderbilt D.
Metric tensor formulation of strain in density-functional perturbation theory.
\textit{Phys Rev B} \textbf{2005};71(3):035117.


\bibitem{Gonze2020}
Gonze X, Amadon B, Antonius G, et al.
The Abinitproject: Impact, environment and recent developments.
\textit{Comput Phys Commun} \textbf{2022};248:107042.

\bibitem{Romero2020}
Romero AH, Allan DC, Amadon B, et al.
ABINIT: Overview and focus on selected capabilities.
\textit{J Chem Phys} \textbf{2020};152(12):124102.

\bibitem{Torrent2008}
Torrent M, Jollet F, Bottin F, et al.
Implementation of the projector augmented-wave method in the ABINIT code: Application to the study of iron under pressure.
\textit{Comput Mater Sci} \textbf{2008};42(2):337--351.

\bibitem{jolletGenerationProjectorAugmentedWave2014}
Jollet F, Torrent M, Holzwarth N.
Generation of Projector Augmented-Wave atomic data: A 71 element validated table in the XML format.
\textit{Comput Phys Commun} \textbf{2014};185(4):1246-1254.

\bibitem{Jain2013}
Jain A, Ong SP, Hautier G.
Commentary: The Materials Project: A materials genome approach to accelerating materials innovation.
\textit{APL Mater} \textbf{2013};1(1):011002.

\bibitem{osti_1193780}
%\textcolor{green}{
%K. Persson, “Materials data on te (sg:152) by materials
%project,” (2016).
%}
Persson K. 
Materials data on te (sg:152) by materials project.
\textit{The Materials Project} \textbf{2016}.


\bibitem{ciesielskiPermittivityGeTe2018}
Ciesielski A, Skowronski L, Pacuski W, Szoplik T.
Permittivity of Ge, Te and Se thin films in the 200–1500 nm spectral range. Predicting the segregation effects in silver.
\textit{Mater Sci Semicond Process} \textbf{2018};81:64-67.

\bibitem{dresselhausGroupTheoryApplication2008a}
Dresselhaus MS, Dresselhaus G, Jorio A.
Group Theory: Application to the Physics of Condensed Matter.
Berlin: Springer-Verlag \textbf{2008}.


\bibitem{pineLinearWaveVectorShifts1969}
Pine AS, Dresselhaus G.
Linear Wave-Vector Shifts in the Raman Spectrum of $\ensuremath{\alpha}$-Quartz and Infrared Optical Activity.
\textit{Phys Rev} \textbf{1969};188(3):1489--1496.

\bibitem{Oishi22}
%\textcolor{green}{E. Oishi, Y. Fujii, A. Koreeda, arXiv:2210.07526 [cond-mat.mtrl-sci] (2022)}
Oishi E, Fujii Y, Koreeda A.
Selective observation of enantiomeric chiral phonons in $\alpha$-quartz.
\textit{arXiv preprint};\textbf{2022}:arXiv:2210.07526.


\bibitem{Oiwa22}
Oiwa R, Kusunose H.
Rotation, Electric-Field Responses, and Absolute Enantioselection in Chiral Crystals.
\textit{Phys Rev Lett} \textbf{2022};129(11):116401.

\bibitem{Yamamoto22}
%\textcolor{green}{
%J. Kishine, H. Kusunose, H. M. Yamamoto, arXiv:2208.06071 [cond-mat.mtrl-sci] (2022)
%}
Kishine J, Kusunose H, Yamamoto HM.
On the Definition of Chirality and Enantioselective Fields.
\textit{Isr J Chem} \textbf{2022};62(11-12):e202200049.

\bibitem{Yariv06}
Yariv A, Yeh P.
\textit{Photonics: Optical Electronics in Modern Communication (6th ed.)} 
New York: Oxford University Press \textbf{2007}.

\bibitem{Ohe}
Ohe K, Shishido H, Kato M, Matsuura H, Togawa Y.
Chirality-Induced Selectivity of Phonon Angular Momenta in Chiral Quartz Crystals.
submitted.


\end{thebibliography}

\newpage

%Tables

\begin{table*}[hbtp]
		\label{table:mesurement_HH_HV}
		\centering
		\begin{tabular}{cccc}
				\hline
			    mode & Sample 1 & Ref. \cite{torrieRamanSpectrumTellurium1970} (RT, &  Ref. \cite{pineRamanSpectraLattice1971} (295 K,  \\
			      &(RT, 785 nm)&  514.5 nm) & 514.5 nm)\\
				\hline \hline
				$\Gamma_3^{(1)}$&93  & 91 & 92   \\
				$\Gamma_1$&121  & 120 & 120 \\
				$\Gamma_3^{(2)}$&141  & 140 & 141 \\
				\hline
		\end{tabular}
		\caption{Stokes Raman shift in Sample 1 of Te (cm$^{-1}$). The peak position is measured by linearly polarized Raman scattering spectoroscopy.}
\end{table*}

\newpage

\begin{table*}[hbtp]
			\label{table:mesurement_RL_LR}
			\centering
			\begin{tabular}{cccc}
				\hline
				peak & Sample 1 & Sample 2 & Ref. \cite{pineRamanSpectraLattice1971} (90 K, \\
				 & (RT, 785 nm) & (RT, 785 nm) &  514.5 nm) \\
				\hline \hline
				$-$93  & 0.8 & 0.8   & - \\
				93  & 0.7 & 0.8  & 0.6  \\
				$-$141  & 0.2 & 0.3   & -  \\
				141  & 0.2 & 0.3 & 0.3  \\
				\hline
			\end{tabular}
			\caption{$\Gamma_3$ mode splitting in Sample 1 and 2 Te  (cm$^{-1}$).}
\end{table*}

\newpage

%Figures

\newpage

\begin{figure*}
\includegraphics{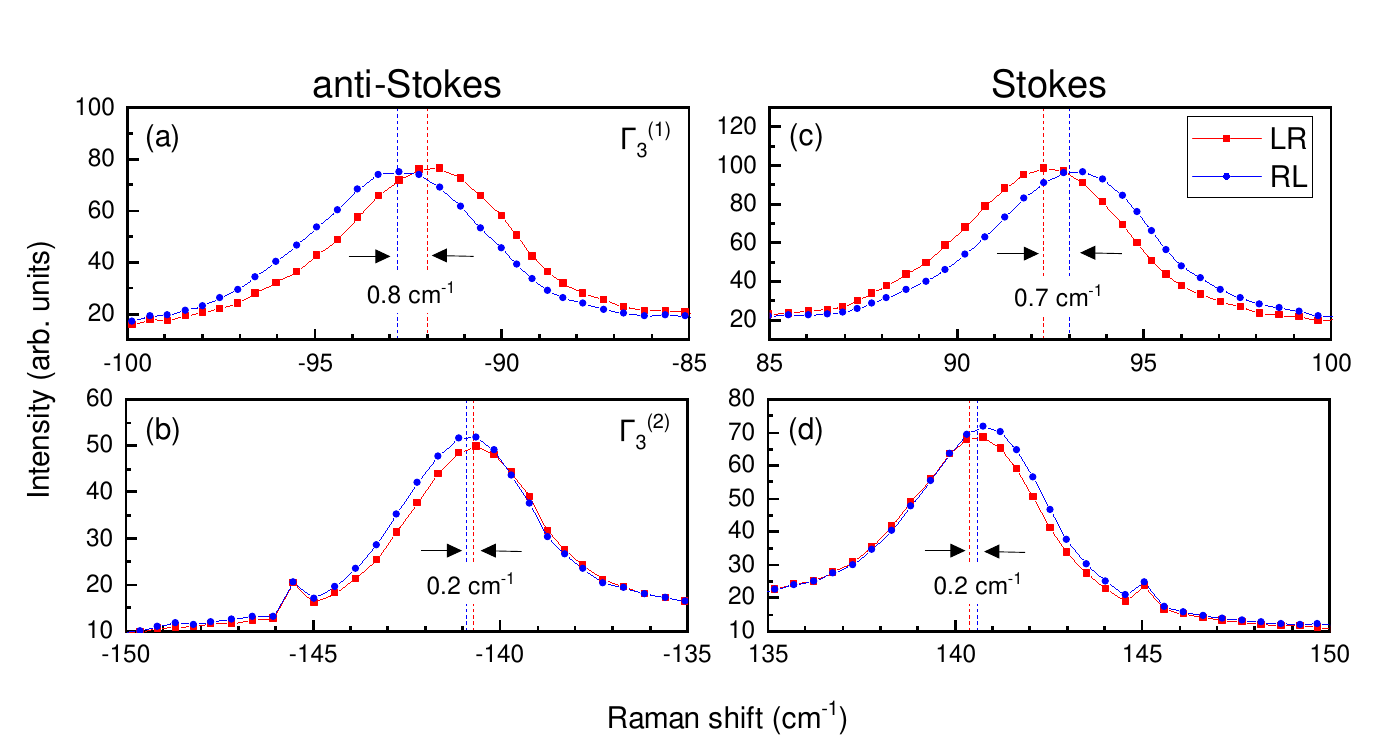}% Here is how to import EPS art
\caption{\label{fig:Te-1266-RL-LR-fixed-WG-RLLR-eng} Raman spectra of Sample 1 Te. Anti-Stokes (a, b) and Stokes (c, d) Raman spectra of the $\Gamma_3^{(1)}$ (a, c) and  $\Gamma_3^{(2)}$ (b, d) modes in the RL (blue line) and LR (red line) configurations and the $\Gamma_3$ doublet splittings (dashed vertical lines) in (a--d).}
\end{figure*}

\begin{figure*}
\includegraphics{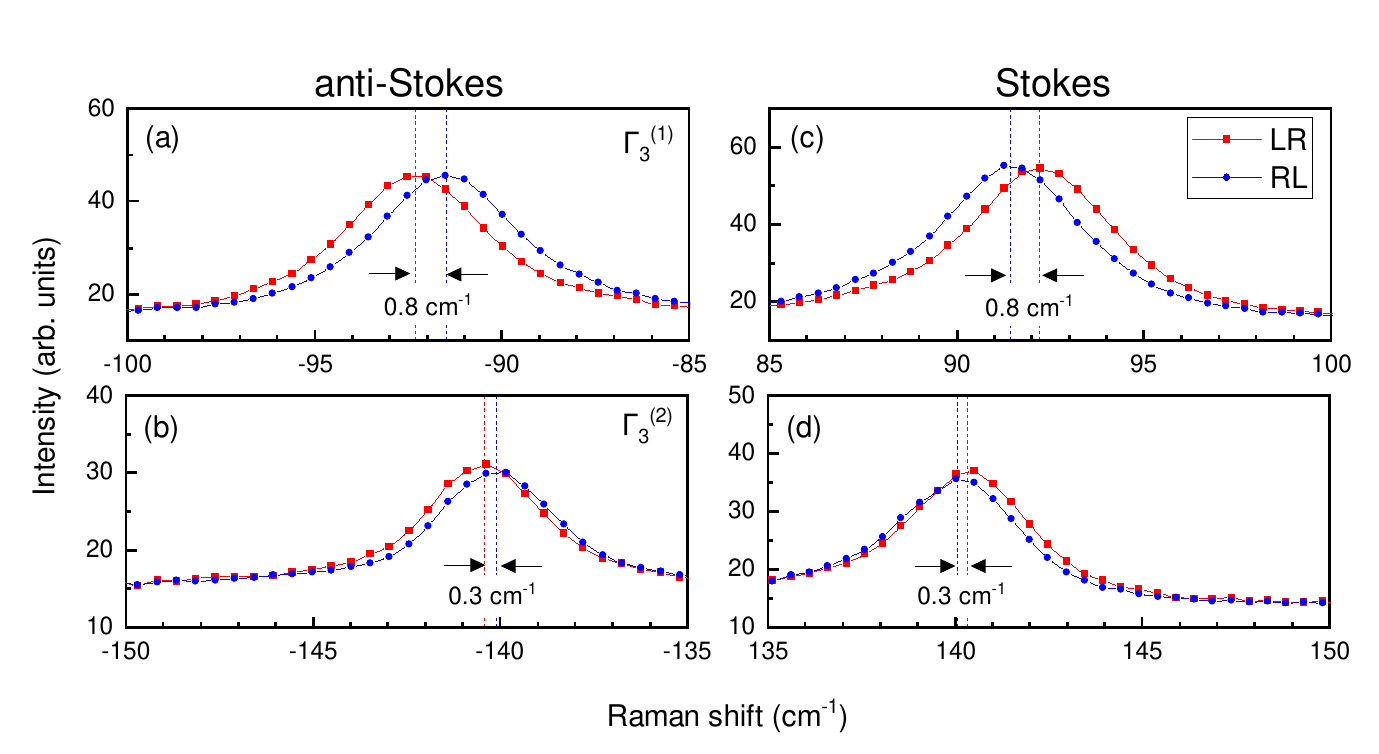}% Here is how to import EPS art
\caption{\label{fig:Te-1266-RL-LR-fixed-WG-RLLR-eng} Raman spectra of Sample 2 Te. Anti-Stokes (a, b) and Stokes (c, d) Raman spectra of the $\Gamma_3^{(1)}$ (a, c) and  $\Gamma_3^{(2)}$ (b, d) modes in the RL (blue line) and LR (red line) configurations and the $\Gamma_3$ doublet splittings (dashed vertical lines) in (a--d).}
\end{figure*}

\newpage

\begin{figure*}
\includegraphics{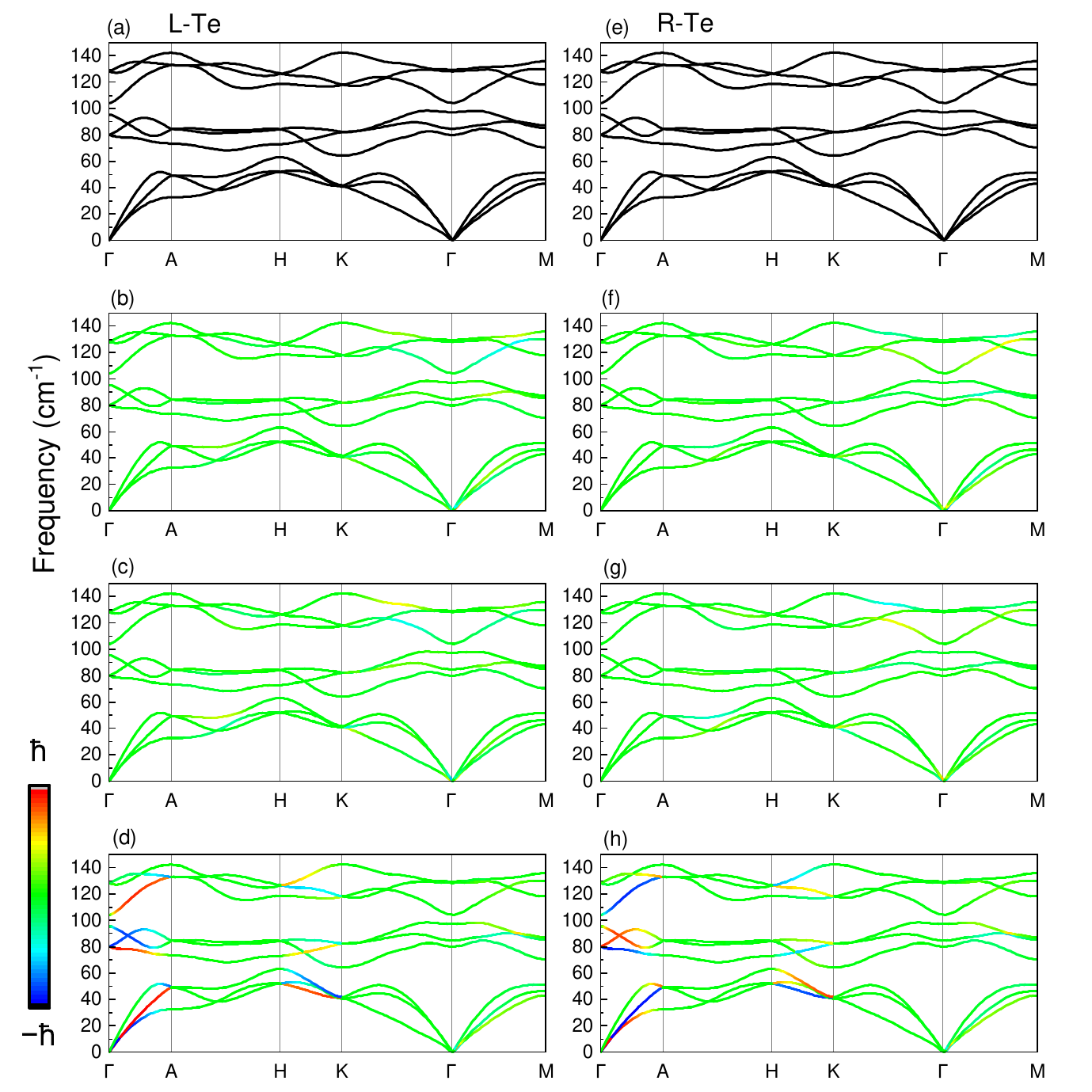}% Here is how to import EPS art
\caption{\label{fig:work_Te_paw_WOSOC_886_25ha_right_AM}\textbf{Left column}
(a) The phonon dispersion curve in the L-Te. (b, c, d) The $x$, $y$, $z$ components of AM in the L-Te, respectively. The red, blue, and green lines indicate the positive, negative, and zero AM, respectively.
\textbf{Right column}
(e) The phonon dispersion curve in the R-Te. (f, g, h) The $x$, $y$, $z$ components of AM in the R-Te, respectively.
}
\end{figure*}

\newpage

\begin{figure*}
\includegraphics{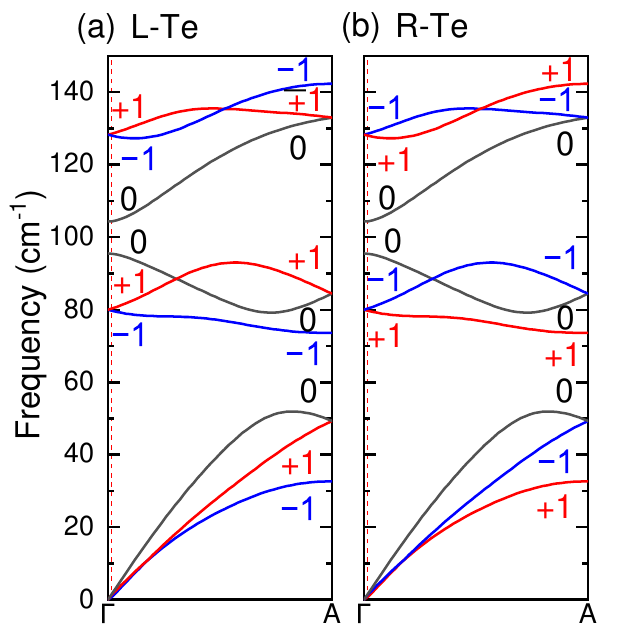}% Here is how to import EPS art
\caption{\label{fig:work_Te_paw_WOSOC_886_25ha_right_left_PAM} (a, b) The spin PAM $m_{\text{PAM}}^{\text{s}}$ in the L- and R-Te, respectively. The red dashed line is the point excited by Raman spectroscopy using a 785 nm laser.}
\end{figure*}

\end{document}